\documentclass[aps,prd,twocolumn,a4paper,superscriptaddress,10pt,showpacs,notitlepage,floatfix]{revtex4-1}

\usepackage{graphicx}

\usepackage{amsmath}
\usepackage{mathtools}
\usepackage{MnSymbol}
\usepackage{dsfont}
\usepackage{bbold}
\usepackage{slashed}

\usepackage{marvosym}
\usepackage[normalem]{ulem}

\usepackage{lipsum}
\usepackage{xcolor}
\usepackage{colortbl}

\usepackage[plainpages=false,pdfpagelabels,pageanchor=false]{hyperref}

\usepackage[squaren]{SIunits}

\def \beq  {\begin{equation}}
\def \eeq  {\end{equation}}

\newcommand{\overbar}[1]{\mkern 1.5mu\overline{\mkern-1.5mu#1\mkern-1.5mu}\mkern 1.5mu}
\newcommand{\MSbar}{\overbar{\text{MS}}}

\newcommand{\Qd}{\!\scriptscriptstyle \mathcal{Q}}

\newcommand{\cI}{\mathcal{I}}
\newcommand{\cR}{\mathcal{R}}
\newcommand{\cQ}{\mathcal{Q}}
\newcommand{\cP}{\mathcal{P}}

\newcommand{\bI}{\mathbf{I}}
\newcommand{\bR}{\mathbf{R}}
\newcommand{\bQ}{\mathbf{Q}}

\newcommand{\Real}{\text{Re}\,}

\DeclareMathOperator*{\res}{\text{Res~~}}

\begin{document}
\allowdisplaybreaks
\begin{abstract}%
Perturbative expansions for short-distance quantities in QCD are factorially divergent and 
this deficiency can be turned into a useful tool to investigate nonperturbative corrections. In this work, we use this approach
to study the structure of power corrections to parton quasi-distributions and 
pseudo-distributions which appear in lattice calculations of parton distribution functions. As the main result, 
we predict the functional dependence of the leading power corrections to {quasi(pseudo)}-distributions on the Bjorken $x$ variable. We also show that these corrections 
can be strongly affected by the normalization procedure.
\end{abstract}

\title{Power corrections and renormalons in parton quasi-distributions}

\author{Vladimir~M.~\surname{Braun}}
\author{Alexey~\surname{Vladimirov}}
\author{Jian-Hui~\surname{Zhang}}

\affiliation{Institut f{\"u}r Theoretische Physik, Universit{\"a}t Regensburg,\\
Universit{\"a}tsstra{\ss}e 31, 93040 Regensburg, Germany}

\maketitle
\section{Introduction}\label{sect_introduction1}%

Lattice calculations in QCD have demonstrated the ability to complement, and in certain cases with the exceeding precision, significant amount of experimental measurements.  Now, the lattice evaluation of parton distribution functions (PDFs) are coming on the agenda. New techniques are being explored aiming at the access of PDFs directly in the momentum fraction space, in addition to the standard approach that allows one to calculate first Mellin moments of PDFs. The existing actual proposals~\cite{Aglietti:1998ur,Abada:2001if,Detmold:2005gg,Braun:2007wv,Ji:2013dva,Ma:2017pxb} differ in details but have a common general scheme: PDFs are extracted from the lattice calculations of suitable Euclidean correlation functions using QCD collinear factorization in the continuum theory. 

A particularly popular suggestion~\cite{Ji:2013dva} that has triggered  a lot of recent activity \cite{Lin:2014zya,Alexandrou:2015rja,Chen:2016utp,Alexandrou:2016jqi,Lin:2017ani,Zhang:2017bzy,Chen:2017gck,Chen:2017mzz,Alexandrou:2017huk, Alexandrou:2018pbm,Chen:2018xof,Chen:2018fwa,Alexandrou:2018eet,
Lin:2018qky}, introduces a concept of a parton \emph{quasi-distribution} (qPDF) defined as a Fourier transform of the nonlocal quark-antiquark operator connected by the Wilson line. QPDF is then matched to PDF either directly in the $\MSbar$ scheme or using the large-momentum factorization scheme at the intermediate step (``Large Momentum Effective Theory'' (LaMET)~\cite{Ji:2014gla,Izubuchi:2018srq}). The latter technique is useful to emphasize that, after the Fourier transform, the hadron momentum $p$ remains to be the only dimensional parameter. So, the relevant scale of the QCD coupling is related to $p$ up to a dimensionless constant and also the higher-twist corrections are generically suppressed by powers of the momentum $p$. A closely related quantity, a \emph{pseudo-distribution} (pPDF) was introduced in~\cite{Radyushkin:2017cyf,Orginos:2017kos,Karpie:2017bzm}.
 
Lattice calculations of PDFs using these new methods are presently moving from exploratory stage towards precision calculations, therefore questions like whether the higher-twist (power suppressed) corrections are well under control have to be addressed. One possibility to investigate the impact of higher-twist corrections is to extract PDFs from the global analysis of many Euclidean correlation functions introducing 
{such} corrections as free parameters. This approach is
showcased in \cite{Bali:2018spj} using the position-space strategy~\cite{Braun:2007wv,Bali:2017gfr} 
and the pion light-cone distribution amplitude (LCDA) as an example.
In this case, the analytic structure of higher-twist effects is well understood, so they could have been modelled by just one free parameter.  
The general situation  may be more complicated. Having in mind the multitude of 
observables that can potentially be employed in lattice calculations, see e.g.~\cite{Ma:2017pxb}, 
it is desirable to have a general method to estimate the corresponding higher-twist corrections and their expected Bjorken-$x$ dependence.   
The purpose of this work is to point out that the problem of power-suppressed contributions for such observables 
can be addressed using the concept of renormalons \cite{Beneke:1998ui,Beneke:2000kc}. In what follows we apply this technique 
to qPDFs and pPDFs.

The renormalon approach to the investigation of power corrections is founded on the fact that operators of different twist mix with each other under renormalization, due to the violation of QCD scale invariance through the running of the coupling constant. In cutoff schemes, this mixing is explicit, whereas in dimensional regularization, it manifests itself in factorial divergence of the perturbative series. Independence of a physical observable on the factorization scale implies intricate cancellations between different twists --- the cancellation of renormalon ambiguities. In turn, the existence of these ambiguities in the leading-twist expressions can be used to estimate the size of power-suppressed corrections. Conceptually, it is similar to the estimation of the accuracy of fixed-order perturbative results by the logarithmic scale dependence. The renormalon approach was used before for the study of Bjorken-$x$ dependence of higher-twist corrections in deep-inelastic scattering \cite{Beneke:1995pq,Dokshitzer:1995qm,Dasgupta:1996hh}, fragmentation functions~\cite{Dasgupta:1996ki,Beneke:1997sr}, pion LCDA~\cite{Braun:2004bu} and transverse-momentum-dependent parton distributions~\cite{Scimemi:2016ffw}.

In order to explain how the concept of renormalons can be used to get insight in the structure of power corrections, let us
consider the usual expression for the quasi-distribution~\cite{Ji:2015qla,Izubuchi:2018srq}, 
\begin{align}
\label{qPDF1}
 \mathcal{Q}(x,p) &= \int_{-1}^1 \!\frac{dy}{|y|} C_{\Qd}(\tfrac{x}{y},xp,\mu_F) q(y,\mu_F) + \frac{1}{p^2}  \mathcal{Q}_4(x,p) + \ldots
\end{align}
where $q(y,\mu_F)$ is the quark PDF, $p$ is the hadron momentum, $x$ refers to the momentum fraction. For brevity we do not show the dependence on the renormalization scale. The factorization scale $\mu_F$ has to be taken of the order of $|x| p$ to avoid large logarithms.  The coefficient function $C(x,p,\mu_F) = \delta(1-x) + \mathcal{O}(\alpha_s)$ is given by the perturbative expansion. The correction of ${\cal O}(\alpha_s)$ was first computed for the flavor-nonsinglet case in \cite{Xiong:2013bka}, see also~\cite{Stewart:2017tvs}. 

To understand the role of renormalons, it is necessary to examine carefully the separation made in~\eqref{qPDF1} between the leading term and the higher twist addenda. Let us assume for a moment that the factorization is done using a hard cutoff $\Lambda_{\rm QCD} \ll \mu_F  \ll p$, i.e. the contributions with loop momenta $|k| > \mu_F$ are included  in the coefficient function, whereas the contributions with $|k| < \mu_F$ are included in PDF. In this scheme, 
the coefficient function has the following expansion at $p\to\infty$
\begin{align}
\label{C-cutoff}
 C_{\Qd}(x,p,\mu_F) &= \delta(1-x) + c_1 \alpha_s + c_2\alpha_s^2 + \ldots 
\notag\\ &\quad{}
- \frac{\mu_F^2}{p^2} D_{\Qd}(x) +\ldots\,,
\end{align}    
where $c_k= c_k(x,\ln p^2/\mu_F^2)$ are the perturbative coefficients depending logarithmically on the scales and the $D_{\Qd}$-term represents the leading power correction. Since the left-hand-side of Eq.~\eqref{qPDF1} does not depend on $\mu_F$, any such dependence should cancel on the right-hand-side. In particular, the logarithmic dependence on the scale in $c_k(x,\ln p^2/\mu_F^2)$ is canceled by the scale-dependence of PDF $q(x,\mu_F)$. The cancellation of the power dependence, on the other hand, must involve the twist-four contribution $\mathcal{Q}_4(x,p)$. Thus, in this factorization scheme one expects that
\begin{align}
\label{Q4-cutoff}
 \mathcal{Q}_4(x,p,\mu_F) = \mu_F^2 \int_{-1}^1 \!\frac{dy}{|y|} D_{\Qd}(\tfrac{x}{y}) q(y,\mu_F) + \widetilde{\mathcal{Q}}_4(x,p,\mu_F)\,,  
\end{align} 
where $\widetilde{\mathcal{Q}}_4$ depends on $\mu_F$ at most logarithmically. 
Appearance of the term $\sim \mu_F^2$ can be traced to quadratic ultraviolet divergence (in addition to the logarithmic ultraviolet divergence) 
of the twist-four operators that are responsible for the power correction, in the cutoff scheme. 
One can prove that the cutoff dependence $\sim \mu_F^2$ of the higher-twist operators is indeed that of Eq.~\eqref{C-cutoff}.    

In practice, perturbative calculations are usually done using dimensional regularization. In this case, power-like terms as in~\eqref{C-cutoff} do not appear. The price to pay is that the coefficients $c_k$ computed in a $\MSbar$ scheme grow factorially with the order $k$. The factorial growth implies that the sum of the perturbative series is only defined to a power accuracy and this ambiguity (renormalon ambiguity) must be compensated by adding a non-perturbative higher--twist correction. The detailed analysis shows~\cite{Beneke:1998ui,Beneke:2000kc} that the divergent large-$k$ behavior (the renomalon) of the coefficients is in the one-to-one correspondence with the sensitivity to extreme (small or large) loop momenta. In particular, infrared renormalons in  the leading-twist coefficient function are compensated by ultraviolet renormalons in the matrix elements of twist-four operators. In this way the same picture as in the cutoff scheme re-appears in dimension regularization.

Returning to \eqref{Q4-cutoff}, we observe that the quadratic term in $\mu_F$ is spurious since its sole purpose is to cancel the similar contribution to the coefficient function. Therefore, it does not contribute to any physical observable. The idea of the renormalon model of the power corrections \cite{Beneke:1995pq,Dokshitzer:1995qm,Dasgupta:1996hh,Dasgupta:1996ki,Beneke:1997sr,Braun:2004bu} is that, with a replacement of $\mu_F$ by a suitable non-perturbative scale, this contribution reflects the order and the functional form of actual power-suppressed contribution. Assuming this ``ultraviolet dominance''~\cite{Braun:1995tp,Beneke:1998ui,Beneke:2000kc} one obtains the following  model:
\begin{equation}
 \mathcal{Q}_4(x,p,\mu_F) = \kappa 
\Lambda_{\rm QCD}^2 \int_{-1}^1 \!\frac{dy}{|y|} D_{\Qd}(\tfrac{x}{y})\, q(y,\mu_F)\,,
\label{UV_Lambda_int}
\end{equation}
with the dimensionless coefficient $\kappa = \mathcal{O}(1)$ which cannot be fixed within theory and remains a free parameter.

In this work we calculate the function $D_{\Qd}(x)$ for different versions of the qPDFs (pPDFs) in the so-called bubble-chain approximation~\cite{Beneke:1998ui}, which is our main result. This calculation reveals that the power corrections to quasi- and pseudo-distributions have the
following generic structure
\begin{align}
 \cQ (x,p) &=  q(x) \Big\{1 + \mathcal{O}\Big( \frac{\Lambda^2}{p^2x^2(1-x)}\Big) \Big\},  
\notag\\
\mathcal{P}(x,z) &= q(x)\Big\{1 + \mathcal{O}\big( z^2\!\Lambda^2{(1-x)}\big)\Big\},
\end{align}
respectively, and can be affected significantly by normalization. In particular, the normalization of the involved matrix elements to their value at zero momentum considerably reduces the power correction for the qPDFs at smaller-$x$ at the cost of a strong enhancement at larger values. We emphasize that the leading power correction to qPDF at $x\to 0$  is enhanced by two powers of the Bjorken $x$ variable. For pPDFs the power corrections are suppresed at $x\to1$, which, unfortunately,  does not hold after the normalization procedure of Ref.~\cite{Orginos:2017kos} (but can be upheld with a different choice). Additionally, as a byproduct of bubble-chain calculation, we have obtained the large-$n_f$ part, $n^k_f \alpha_s^{k+1}$, of the
leading twist coefficient function to all orders in perturbation theory.

The presentation is organized as follows. In Sect.~2 we formulate
our program in more precise terms. We define qPDFs and  pPDFs as 
particular Fourier transforms of the position space (Ioffe-time) distributions, 
discuss briefly the light-ray operator product expansion (OPE) and the target mass corrections, and  
introduce the relevant techniques (Borel transform) and the systematic approximation
(large-$n_f$ expansion) that will be used throughout the rest of
the work. In Sect. 3 we present our result for the Borel transform of the leading-twist
coefficient function and discuss the structure of its singularities. 
The leading power corrections to various versions of the quasi-distributions are obtained in Sect.~4. 
We collect there the relevant analytic expressions and also present the results of a numerical 
study using realistic parametrizations for the valence quark PDFs. 
The final Sect.~5 is reserved for the summary and conclusions.

\section{General formalism}\label{sect_formalism}%
\subsection{Parton quasi-distributions}\label{subsect_definitions}%
Let us start with the following nucleon matrix element
\begin{align}
\label{def:ME}
 &\langle N(p)| \bar q (zv) \gamma_\alpha[zv,0] q(0) |N(p)\rangle  = 
\phantom{\frac{2v_\alpha}{v^2} (pv) \mathcal{I}^\parallel (v^2z^2\!,pvz,\mu)}
\notag\\
&\hspace*{2.0cm} = \frac{2v_\alpha}{v^2} (pv)\, \mathcal{I}^\parallel (v^2z^2\!,pvz,\mu)
\notag\\
&\hspace*{2.0cm}\quad+ 2\left(p_\alpha- \frac{v_\alpha}{v^2} (pv) \right) \mathcal{I}^\perp (v^2z^2\!,pvz,\mu)\,,
\end{align}
where
\begin{align}
\label{def:Wilson}
 [zv, 0] &= \text{Pexp}\Big[i g \int_0^z du\,  v^\mu A_\mu(uv)\Big]\,,
\end{align}
$p^\mu$ is the nucleon momentum, $p^2 = m^2$, $v^\mu$ is a given four-vector and $z$ a real number. All notations correspond to Minkowski space. In the following, we keep the normalization of the four-vector $v_\mu$ arbitrary, keeping in mind that $v^2 <0$ in the lattice calculation. We suppress flavor indices and tacitly assume considering the flavor-non-singlet combination of the matrix elements in what follows. We also neglect quark masses. 

The operator product in Eq.\,\eqref{def:ME} suffers from ultraviolet (UV) divergences and has to be renormalized. The argument $\mu$ of matrix elements $\mathcal{I}$ refers to the renormalization scale. The renormalization-scale dependence can be studied by going over to an effective theory~\cite{Craigie:1980qs,Dorn:1986dt,Ji:2017oey} such that the Wilson line is substituted by the propagator of an auxiliary field. For time-like separations the resulting theory is  the heavy-quark effective theory (HQET) and the renormalization factor of interest is the renormalization factor (squared) for the heavy-light quark current, see e.g.~\cite{Chetyrkin:2003vi}. We are not aware of a calculation for such a renormalization factor at space-like separations beyond one-loop order, however, to this accuracy there is no difference from the time-like case.

The ``longitudinal'' and ``transverse'' invariant functions in Eq.~\eqref{def:ME} (with respect to $v_\mu$) correspond to particular projections of the matrix element that are employed in lattice calculations: 
\begin{align}\label{invfuncdef}
 \langle N(p)| \bar q (zv) \slashed{v} [zv,0]q(0) |N(p)\rangle  & = 2 (pv) \mathcal{I}^\parallel (v^2z^2,pvz,\mu)\,,
\notag\\
  \langle N(p)| \bar q (zv) \slashed{\epsilon}[zv,0] q(0) |N(p)\rangle &= 2 (p\epsilon) \mathcal{I}^\perp (v^2z^2,pvz,\mu)\,,
\end{align}
where $(\epsilon\cdot v) =0$. Eq.~(\ref{invfuncdef}) is the starting point for the construction of quasi-distributions.
 
The invariant functions  $\mathcal{I}^{\parallel}$, $\mathcal{I}^{\perp}$ \eqref{def:ME}  coincide at the tree level.  Assuming the power counting \begin{align}
\label{power-counting}
  z = \mathcal{O}(\eta)\,, \qquad p = \mathcal{O}(\eta^{-1})\,,\qquad \eta \to 0\,, 
\end{align} 
they can be written in terms of the position-space quark PDF~\cite{Ioffe:1969kf,Braun:1994jq,Hoyer:1996nr}:
\begin{align}
  \mathcal{I}^{\parallel(\perp)}(z^2v^2,pvz, \mu\sim 1/|vz|) &=  I(pvz, \mu_F \sim 1/|vz|)
\notag\\&\quad
 +\mathcal{O}(\alpha_s, \eta^2)\,, 
\end{align}
where  $\mu_F$ is the factorization scale, and 
\begin{align}
\label{ITD}
   I(pvz,\mu_F) &=  \int_{-1}^1 dx \, e^{ixz\,pv} q(x,\mu_F)\,.
\end{align} 
Here $q(x,\mu_F)$ for $x>0$ is the quark PDF, and for $x<0$ is the antiquark PDF. The position-space PDFs $I$ are known as the Ioffe-time distribution (ITD). To distinguish ITD $I$ from the functions $\mathcal{I}$, we refer to the functions $\mathcal{I}^{\parallel}$, $\mathcal{I}^{\perp}$  as the Ioffe-time \emph{quasi}-distributions (qITDs), following the terminology introduced in \cite{Ji:2013dva}.
 
The qPDFs $\mathcal{Q}^{\parallel},\mathcal{Q}^{\perp}$~\cite{Ji:2013dva} and the pPDF $\mathcal{P}$~\cite{Radyushkin:2017cyf}
are defined in terms of the qITDs by the Fourier transform
\begin{align}
\label{def:qPDF+pPDF}
\mathcal{Q}^{\parallel(\perp)}(x,p,\mu) &= |(pv)|\! \int\limits_{-\infty}^\infty \frac{d z }{2\pi} e^{-ix z (pv)} \, \mathcal{I}^{\parallel(\perp)}(z^2v^2, pvz,\mu )\,,
\notag\\
\mathcal{P}(x,z,\mu) &= |z| \int\limits_{-\infty}^\infty \frac{d (pv)}{2\pi} e^{-ix z (pv) } \,  \mathcal{I}^{\perp}(z^2v^2,pvz,\mu)\,.     
\end{align}
In what follows we will tacitly assume that $(pv)>0$ in the discussion of qPDFs and $z>0$ for pPDFs and drop the 
absolute value sign, for brevity.

In renormalization schemes with an explicit regularization scale, the Wilson line in Eq.\,\eqref{def:ME} suffers 
from an additional linear UV divergence that has to be removed. In dimensional regularization, this UV linear divergence reveals itself as a factorial growth of high orders of perturbative series~\cite{Beneke:1994sw}. The linear UV divergence can be removed by a mass renormalization associated with the Wilson line~\cite{Dotsenko:1979wb,Craigie:1980qs,Dorn:1986dt} or by a regularization-independent renormalization~\cite{Martinelli:1995vj,Stewart:2017tvs,Chen:2017mzz}. Given the multiplicative renormalizability of the quasi-PDF operator~\cite{Ji:2017oey,Ishikawa:2017faj,Green:2017xeu}, it can also be removed by considering a suitable ratio of matrix elements involving the same operator, e.g. by normalizing to the same matrix element at zero proton momentum~\cite{Orginos:2017kos}, 
\begin{align}
\label{n-qITD}
     \mathbf{I}(v^2z^2, pvz) =  \mathcal{I}(v^2z^2, pvz, \mu) / \mathcal{I}(v^2z^2, 0, \mu)\,, 
\end{align}
or, alternatively, to the vacuum expectation value
\begin{align}
\label{hat-qITD}
     \widehat{\mathbf{I}}(v^2z^2, pvz) =  \mathcal{I}(v^2z^2, pvz, \mu) / \mathcal{N}(v^2z^2,\mu)\,, 
\end{align}
where
\begin{align}
  \mathcal{N}(v^2z^2,\mu) =  \left(\frac{2 i N_c}{\pi^2 z^3 v^2}\right)^{-1} \langle 0| \bar q(zv) \slashed{v} [zv,0] q(0) |0\rangle\,.
\end{align}
Eqs.~(\ref{n-qITD}) and (\ref{hat-qITD}) will be our main focus in the present paper. By forming the ratio, the scale dependence cancels out (including the usual logarithmic renormalization) and one can define the scale-independent qPDF/pPDF
\begin{align}
\mathbf{Q}(x,p) &= |(pv)| \int_{-\infty}^\infty \frac{d z }{2\pi} e^{-ix z (pv)} \, \mathbf{I}(z, pv)\,,
\notag\\
\mathbf{P}(x,z) &= |z| \int_{-\infty}^\infty \frac{d (pv)}{2\pi} e^{-ix z (pv) } \,  \mathbf{I}(z,pv)\,,     
\end{align} 
and similarly for $\widehat{\mathbf{Q}}(x,p)$ and $\widehat{\mathbf{P}}(x,z)$.
The difference between these two options in the present context is that the normalization to the vacuum correlator 
does not affect the leading $\mathcal{O}(v^2z^2)$ power corrections that are subject of this work (since there is no gauge-invariant operator),
whereas the normalization to the value at zero momentum, as we will see, has a substantial effect. 

%
\subsection{Light-ray OPE and target mass corrections}\label{subsect_OPE}%

The general approach to collinear factorization of QCD amplitudes in the position space is provided by the light-ray OPE~\cite{Anikin:1978tj,Anikin:1979kq,Balitsky:1987bk,Mueller:1998fv,Balitsky:1990ck,Geyer:1999uq}. For illustration consider the ``longitudinal'' projection. Specializing to the present case (forward matrix elements) we write
\begin{align}
\label{light-ray-OPE}
 \bar q(zv) \slashed{v} [zv,0] q(0) &= \int\limits_0^1\!d\alpha \, H^\parallel(z,\alpha,\mu_F)
\notag\\&\quad \times \Pi_{\rm l.t.}^{\mu_F}[\bar q (\alpha z v) \slashed{z} q(0)] + \ldots,
\end{align} 
where we include the renormalization factor in the coefficient function $H^\parallel = \delta(1-\alpha) + \mathcal{O}(\alpha_s)$ and set 
the renormalization scale $\mu$ to be equal to the factorization scale $\mu_F$. 
Finally,  $\Pi_{\rm l.t.}^{\mu_F}[\ldots]$ is the leading-twist projection operator defined below and
ellipses stand for the higher-twist contributions.

The leading-twist projection of the nonlocal quark-antiquark operator is 
defined as the generating function of \emph{renormalized} 
local leading-twist operators  (traceless and symmetrized over all indices)
\begin{align}
\Pi_{\rm l.t.}^{\mu_F}[\bar q (z v) \slashed{v} q(0)]
&= \sum_{n=1}^\infty \frac{z^{n-1}}{(n-1)!} v^{\mu_1}\ldots v^{\mu_n} O_{\mu_1\ldots\mu_n}^{n}(0)\,,
\label{eq:LTprojector}
\end{align}
where 
\begin{align}
  O_{\mu_1\ldots\mu_n}^{n}(z) &= 
\bar q(0)\gamma_{(\mu_1}\!\!\stackrel{\leftarrow}{D}_{\mu_2}\ldots \!\stackrel{\leftarrow}{D}_{\mu_{n})} q(0)\,.  
\end{align}
Here we indicate trace subtraction and symmetrization by enclosing the involved Lorentz indices in parentheses,
for example $O_{(\mu\nu)} = \frac12 (O_{\mu\nu}+ O_{\nu\mu}) - \frac14 g_{\mu\nu} O_{\lambda}^{~\lambda}$.

The light-ray OPE differs from the usual short-distance Wilson  expansion in local operators by imposing a different power counting. In the short-distance expansion one assumes that the distance between the quarks is small, $|z| \sim \eta \Lambda^{-1}_{\rm QCD}$ with $\eta\to 0$, and the operator matrix elements are of order unity in this limit, $\langle O_{\mu_1\ldots\mu_n}^{n} \rangle \sim \Lambda^{n}_{\rm QCD}$. In this case only a finite number of terms contribute to the r.h.s. of Eq.~\eqref{eq:LTprojector}, whereas the rest as well as the higher-twist operators must be added to 
higher orders of OPE, starting  from $\mathcal{O}(\eta^2)$. In other words, the relevant expansion parameter is the mass-dimension of an operator. 
The light-ray OPE assumes instead that the leading-twist operators scale as $\langle O_{\mu_1\ldots\mu_n}^{n} \rangle \sim \eta^{-n} \Lambda^{n}_{\rm QCD}$ so that $ z^n v^{\mu_1}\ldots v^{\mu_n} \langle O_{\mu_1\ldots\mu_n}^{n} \rangle  = \mathcal{O}(1)$. In this case, the series in~\eqref{eq:LTprojector} must be resummed to all orders, revealing its non-local "light-ray" structure. For a generic hadronic matrix element of leading twist operators one has
\begin{align}
 \langle p | O_{\mu_1\ldots\mu_n}^{n} |p\rangle \sim p_{(\mu_1}\ldots p_{\mu_n)} \langle\!\langle O^{n} \rangle\!\rangle\,, 
\end{align}      
where the reduced matrix element $\langle\!\langle O^{n} \rangle\!\rangle = \mathcal{O}(1)$. Therefore, the light-ray OPE is an adequate approximation if the hadron has large momentum, $|pv| = \mathcal{O}(\eta^{-1})$ and hence $ z\,pv =  \mathcal{O}(1)$. Higher-twist operators of the same dimension have smaller spin (by definition) and as a consequence, their matrix elements are power-suppressed \footnote{Strictly speaking, the expansion in powers of the large momentum corresponds to the classification in terms of the so-called collinear twist, see e.g.~\cite{Geyer:1999uq,Geyer:2000ig}.}.  

Note that the above power counting is applicable both in Minkowski and Euclidean space. In Minkowski space, one can 
go over to a different reference frame where all components of the momentum are of order $\Lambda_{\rm QCD}$ 
and simultaneously the separation between the quarks is almost light-like, $|v_\mu| = \mathcal{O}(1)$ but 
$v^2 = \mathcal{O}(\eta^2) \to 0$. 

On the calculation level, the light-ray OPE provides one with a convenient framework to operate with the leading-twist projected operators~\eqref{eq:LTprojector} avoiding the local expansion. Light-ray operators can be viewed as analytic operator functions of the separation 
between the quarks (all short-distance and light-cone singularities are subtracted). They satisfy the Laplace 
equation~\cite{Balitsky:1987bk}
\begin{align}
 \frac{\partial}{\partial v^\mu} \frac{\partial}{\partial v_\mu}  \Pi_{\rm l.t.}^{\mu_F}[\bar q (z v ) \slashed{v} q(0)] =0\,
\end{align}
with the boundary condition $\Pi_{\rm l.t.}^{\mu_F}\to 1 $ at $v^2 \to 0$. 
Explicit expressions for the projection operator $\Pi_{\rm l.t.}^{\mu_F}$ can be 
found in~\cite{Balitsky:1987bk,Geyer:1999uq,Geyer:2000ig,Braun:2011dg}. 
The light-ray OPE combined with the background field method is the standard technique, e.g., 
in light-cone sum rules~\cite{Balitsky:1989ry}, where it is used for the calculation of higher-twist contributions, 
and for the derivation of the evolution equations for off-forward parton distributions~\cite{Belitsky:1999hf,Belitsky:1998gc}. 
The renormalization group kernel for the evolution of flavor-singlet light-ray operators in the general off-forward kinematics 
is known to three-loop accuracy~\cite{Braun:2017cih}. 

The nucleon matrix element of the leading-twist projected operator~\eqref{eq:LTprojector} defines the leading-twist 
quark PDF,
\begin{eqnarray}
\lefteqn{ \langle N(p)| \Pi_{\rm l.t.}^{\mu_F}[\bar q (zv) \slashed{v} q(0)] |N(p)\rangle}
\notag\\ 
&=& 2 \int_{-1}^1\!dx\, \Pi_{\rm l.t.}[(p\cdot v) e^{ i x z (pv)}] q(x,\mu_F)\,, 
\end{eqnarray}
where~\cite{Balitsky:1990ck} 
\begin{align}
\label{Pi-exp}
\Pi_{\rm l.t.}[(pv) e^{ i x z(pv)}]
&= (pv)\Big[1 - \frac{1}{4} \frac{m^2 v^2}{(pv)^2} z\frac{d}{dz} \Big] e^{ ixz(pv)}
\nonumber\\&\quad
+ \mathcal{O}(m^4)\,. 
\end{align}
The second term in the square brackets is the leading nucleon mass correction,  which is the position-space counterpart of the Nachtmann target mass correction~\cite{Nachtmann:1973mr} in the deep-inelastic scattering (DIS). The all-order expression in powers of the nucleon mass for the leading-twist projected exponential function can be found in~\cite{Balitsky:1990ck}.

Taking the nucleon matrix element of the operator relation in Eq.~\eqref{light-ray-OPE} we obtain the 
factorization theorem for the ``longitudinal'' qITD in the form
\begin{align}
\mathcal{I}^\parallel   = &
 \int\limits_0^1 d\alpha \, H^\parallel(z,\alpha,\mu_F)
  \Big[1 - \frac{1}{4} \frac{m^2 v^2}{(pv)^2} z\frac{d}{dz} +\ldots \Big]\\ & \nonumber
\times I(\alpha z pv , \mu_F)
+\ldots\,,
\end{align}
where $I(\alpha z pv,\mu_F)$ is  ITD~\eqref{ITD} and ellipses stand for the higher-power target mass 
and ``genuine'' higher-twist corrections.

Making the Fourier transformation \eqref{def:qPDF+pPDF} one obtains, to the leading twist accuracy,
the factorization theorem for the ``longitudinal'' qPDF,
\begin{align}
\label{qPDF2}
 \mathcal{Q}^\parallel(x,p) &= \int_{-1}^1 \!\frac{dy}{|y|} C^\parallel_{\Qd}(\tfrac{x}{y},xp,\mu_F)
\Big[1 + \frac{1}{4} \frac{m^2 v^2}{(pv)^2} \frac{d}{dy} y +\ldots \Big] 
\notag\\&\quad 
\times  q(y,\mu_F)\,, 
\end{align}
where the coefficient function is given by 
\begin{align}
  C^\parallel_{\Qd}(\tfrac{x}{y},xp,\mu_F) &= (pv)|y|\!\int\limits_{-\infty}^{\infty}\!\frac{dz}{2\pi}\! \int\limits_0^1\!\! d\alpha\,
e^{i(pv)z(x-\alpha y)}
\notag\\&\quad \times H^\parallel(z,\alpha,\mu_F)\,. 
\end{align} 
In particular, at the leading order
\begin{align}
\label{target-parallel}
 \mathcal{Q}^\parallel(x,p) &= q(x) + \frac{1}{4} \frac{m^2 v^2}{(pv)^2} \big[x q'(x)  + q(x)\big]
+ \mathcal{O}\left({m^4}/{p^4}\right),
\end{align}
where $q'(x) = (d/dx) q(x)$. Note that the derivative applied to the quark PDF lowers the power $q(x) \stackrel{x \to 1}{\sim} (1-x)^p$ by one unit so that the mass correction is effectively enhanced by the factor $1/(1-x)$ as compared to the quark distribution itself at large Bjorken $x$. The similar enhancement of the target mass correction at $x\to 1$ is familiar from DIS~\cite{Nachtmann:1973mr}.

The target mass correction for the ``transverse'' qPDF can be calculated in a similar way,
starting from the nonlocal operator with an open Lorentz index \eqref{def:ME} and 
using the operator identity~\cite{Balitsky:1987bk}
\begin{align}
\Pi_{\rm l.t.}^{\mu_F}[\bar q (z v) \gamma_\alpha q(0)]
=  \int\limits_0^\infty \!dt\,\frac{\partial}{\partial v^\alpha} 
\Pi_{\rm l.t.}^{\mu_F}[\bar q ( t z v) \slashed{v} q(0)]\,,
\end{align}   
where the Wilson line between the quarks is implied. One obtains, at the tree level,
\begin{align}
\label{target-perp}
 \mathcal{Q}^\perp(x,p)  &=  q(x) + \frac{1}{4} \frac{ m^2 v^2 }{(pv)^2} \Big[ x q'(x) + 3 q(x) \Big]
\notag\\&\quad
-  \frac{1}{2} \frac{m^2 v^2}{(pv)^2} \theta(|x|<1)\!\! \int_{|x|}^1\! \frac{dy}{y} q(x/y) 
+ \mathcal{O}\left({m^4}/{p^4}\right),
\end{align}
and the target mass correction to the pPDF
\begin{align}
\label{target-pPDF}
 \mathcal{P}(x,z) &=  q(x) +  \frac14 z^2 v^2 m^2 x^2 \theta(|x|<1) \int_{|x|}^1 \frac{dy}{y} q(x/y)
\notag\\&\quad
+ \mathcal{O}\left({m^4}/{p^4}\right)\,.
\end{align}
The target mass corrections in Eqs.~(\ref{target-parallel}) and (\ref{target-perp}) agree with those derived in Ref.~\cite{Chen:2016utp} when expanded to $\mathcal O(m^2v^2/(pv)^2)$. Interestingly, the target mass correction to the pPDF is suppressed as $\mathcal{O}(1-x)$ at $x\to 1$ and not enhanced $\mathcal{O}(1/(1-x))$ 
in contrast to that to the qPDFs (\ref{target-parallel}) and the structure functions in DIS.

\subsection{Borel transform and renormalons}

The coefficient function $H^\parallel$ in Eq.~\eqref{light-ray-OPE} and the similar coefficient function 
$H^\perp$ in the $\MSbar$ scheme have the perturbative expansion 
\begin{align}
 H = \delta(1-\alpha) + \sum_{k=0}^\infty h_k a_s^{k+1}\,,\qquad a_s = \frac{\alpha_s(\mu)}{4\pi}\,, 
\end{align}
with factorially growing coefficients $h_k\sim k!$.

A convenient way to handle such a series is to consider the Borel transform
\begin{align}
  B[H] (w) = \sum_{k=0}^\infty \frac{h_k}{k!}\left(\frac{w}{\beta_0}\right)^k 
\end{align}
where powers of $\beta_0 = 11/3 N_C -2/3 n_f$ are inserted for the later convenience. The Borel image can be used as a generating function for the fixed-order coefficients
\begin{align}
 h_k = \beta_0^k \left(\frac{d}{dw}\right)^k  B[H] (w)\big|_{w=0}\,.
\end{align}
Moreover, the sum of the series can be obtained as the integral over positive values of the Borel parameter $w$
\begin{align}
\label{BorelIntegral}
H = \delta(1-\alpha)  + \frac{1}{\beta_0} \int_0^\infty \!dw\, e^{-w/(\beta_0 a_s)} B[H](w)\,. 
\end{align}
As it stands, the integral is not defined because the Borel transform generally has singularities on the integration path, known
as (infrared) renormalons. One can adopt a definition of the integral deforming the contour above or below the real
axis, or as the principle value. These definitions are arbitrary, and their difference, which is exponentially small in the 
coupling, must be viewed as an intrinsic uncertainty of perturbation theory that has to be removed by adding 
power-suppressed nonperturbative corrections. Another potential problem concerns the convergence of the Borel integral at 
$w\to\infty$. Since the quantity of interest depends on the single hard scale $1/|z v|$, dimension counting requires that the Borel
transform can be written as $(\mu^2 z^2 v^2 )^w$ times a function $F(w)$ of the Borel parameter and dimensionless kinematic variables.
Combining $(\mu^2 z^2 v^2)^w$ with $e^{-w/(\beta_0 a_s(\mu))}$ one sees that the (principal value) integral is convergent, provided 
the distance between the quarks $|zv|$ is sufficiently small compared to $1/\Lambda_{\rm QCD}$ and $F(w)$ does not increase exponentially
at $w\to\infty$, which is the case of all known examples. 

Naturally, a full all-order calculation cannot be performed. Instead, we employ the approximation~\cite{Beneke:1998ui,Beneke:2000kc}   
restricting ourselves to the perturbative series generated by the running-coupling effects in the one-loop diagrams, i.e. 
using QCD coupling at the scale of the gluon virtuality. 
Such contributions can be traced by computing  
the diagrams with the insertion of $k$ fermion loops in the one-loop diagram
and replacing  $- \frac23 n_f \mapsto   \beta_0 = \frac{11}{3} N_c - \frac23 n_f$, see Appendix A.
Another, equivalent technique \cite{Ball:1995ni} is based on the calculation of one-loop diagrams with an effective gluon mass.

\begin{figure*}[tbp]
\centering
 \includegraphics[width=0.95\textwidth]{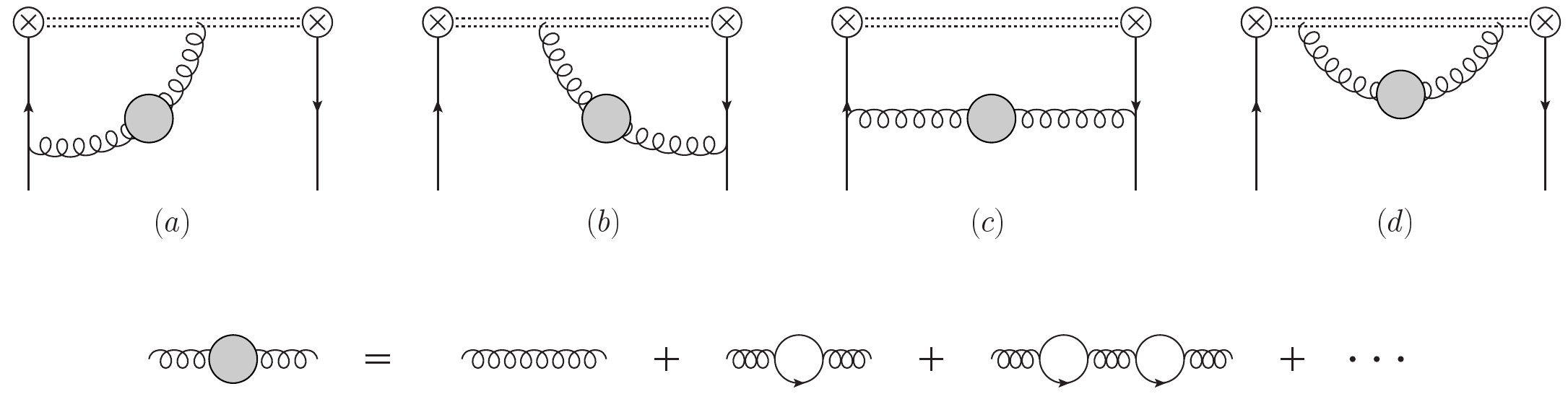}
\caption{Bubble-chain contribution to the coefficient function. The Wilson 
line factor is shown by the double dotted line.}
\label{fig:bubblechain}
\end{figure*}

The singularity structure of the Borel transform can be extracted separately without explicit evaluation of the bubble-chain. It can be done by replacing the running coupling constant in the loop diagrams by its effective form,
\begin{align}
  \beta_0 a_s(-k^2) =  \int_0^\infty \!dw\, e^{\frac53 w} \left(\frac{\Lambda^2}{-k^2}\right)^w ,
\end{align}
where $\Lambda \equiv \Lambda_{\rm QCD}^{\MSbar}$ and the factor $e^{\frac53 w}$ represents the $\MSbar$ scheme. Such replacement leads to the modified form of the gluon propagator 
\begin{align}\label{modgluonprop}
  \frac{1}{-k^2-i\epsilon} ~\mapsto~ \frac{(\Lambda^2)^w}{(-k^2-i\epsilon)^{1+w}}\,.
\end{align}
At one-loop level, the pole structure in $w$ reproduces the pole structure for the Borel plane. In our work, we have used both methods of calculation for the cross-check of the result.

%
\section{Large-$\beta_0$ coefficient functions}\label{sect_bubbles}%
%

The leading contributions to the renormalon singularities in the coefficient functions for qITDs are shown in Fig.~\ref{fig:bubblechain},
see Appendix A for technical details. We obtain   
\begin{align}
\label{B[H]}
B[H^\parallel](w)&=\frac{2C_F}{w} \biggl\{\Big[\frac{1+\alpha^2}{1-\alpha}
- \Big(2\alpha\,{}_2F_1(1,2-w,2+w,\alpha)
\notag\\&\quad
 +\bar \alpha (1-w^2)\Big)\alpha^w\,h_0(w,X)\Big]_+
\notag\\ &\quad
+\delta(\bar \alpha)\bigg[\frac{3(w^2-w-1)}{(w+2)(2w-1)}h_0(w,X)-\frac{3}{2}\bigg]\biggr\}
\notag\\[1mm] &\quad
+\widetilde{R}(w),
\\
B[H^\perp](w)&= B[H^\parallel](w)-4C_F \bar \alpha(1+w)\alpha^w\,h_0(w,X)\,,
\end{align}
where $\bar\alpha = 1-\alpha$,
\begin{align}
& h_0(w,X)=X^w \frac{\Gamma(1-w)}{\Gamma(2+w)},
\notag\\
&X=- \frac{v^2 z^2 \mu^2 e^{5/3}}{4},
\end{align}
and the function $\widetilde{R}(w)$ is defined as the series expansion in terms of another function
\begin{align}
R(w)&=2C_F \biggl\{\Big[\frac{1+\alpha^2}{1-\alpha}\frac{\alpha^wG_0(w)\!-\!1}{w}+\alpha^w \bar \alpha (2\!+\!w)G_0(w)\Big]_+
\notag\\&\quad
+\frac{\delta(\bar \alpha)}{w}\bigg[\frac{3}{2}-\frac{2w+3}{(w+2)(w+1)}G_0(w)\bigg]\biggr\},
\notag\\
G_0(w)&=\frac{\Gamma(4+2w)}{6\Gamma(1-w)\Gamma(1+w)\Gamma^2(2+w)},
\end{align}
such that
\begin{align}
  R(w) = \sum_n w^n R_n\,, \qquad \widetilde{R}(w) = \sum_n \frac{w^n}{(n+1)!} R_n \,.
\end{align}
The Taylor expansion of the Borel transform at $w=0$ gives the perturbative expansion for the coefficient functions in terms of the coupling constant. 
The $\mathcal{O}(\alpha_s)$ term is
\begin{widetext}
\begin{eqnarray}
\label{NLO}
H^{(1)\parallel}(\alpha)&=&2C_F\Bigg[
\bigg(-\mathbf{L}_\mu \frac{1+\alpha^2}{1-\alpha}+\frac{3-8\alpha+3\alpha^2-4\ln\bar \alpha}{1-\alpha}\bigg)_+
+\delta(\bar \alpha)\bigg(\frac{3}{2}\mathbf{L}_\mu+\frac{7}{2}\bigg)\Bigg],
\\
H^{(1)\perp}(\alpha)&=&2C_F\Bigg[
 \bigg(-\mathbf{L}_\mu \frac{1+\alpha^2}{1-\alpha}+\frac{1-4\alpha+\alpha^2-4\ln\bar \alpha}{1-\alpha}\bigg)_+
 +\delta(\bar \alpha)\bigg(\frac{3}{2}\mathbf{L}_\mu+\frac{5}{2}\bigg)\Bigg],
\end{eqnarray}
and the $n_f$-part of the $\mathcal{O}(\alpha_s^2)$ correction reads 
\begin{eqnarray}
\label{NNLO}
H^{(2)\parallel}(\alpha)&=&
-\frac{4C_Fn_f}{3}\biggl\{-\frac{\mathbf{L}_\mu^2}{2}\Big(\frac{1+\alpha^2}{1-\alpha}\Big)_+-\mathbf{L}_\mu \Big[\frac{1+\alpha^2}{1-\alpha}\Big(\ln\alpha+\frac{2}{3}\Big)+4\frac{\alpha+\ln\bar \alpha}{1-\alpha}\Big]_+
+\delta(\bar \alpha)\Big(\frac{3\mathbf{L}_\mu^2}{4}+\frac{19}{4}\mathbf{L}_\mu+\frac{159}{16}\Big)
\\\nonumber && \qquad+\Big[\frac{1+\alpha^2}{1-\alpha}\Big(\frac{41}{18}+\frac{7}{6}\ln\alpha-\frac{\ln^2\alpha}{4}\Big)
-\frac{4}{1-\alpha}\Big(\text{Li}_2(\alpha)+\ln^2\bar \alpha+\frac{8}{3}\ln\bar \alpha+\ln \bar \alpha \ln\alpha\Big)
-\frac{\alpha(13+6\ln \alpha)}{1-\alpha}\Big]_+ \biggr\}.
\\
H^{(2)\perp}(\alpha)&=&H^{(1)\parallel}(\alpha)+\frac{4C_Fn_f}{3}\bar \alpha\Big(2\mathbf{L}_\mu+2\ln\alpha+\frac{10}{3}\Big),
\end{eqnarray}
\end{widetext}
where
\begin{eqnarray}
\mathbf{L}_\mu=\ln\bigg(\frac{-v^2z^2\mu^2}{4e^{-2\gamma_E}}\bigg).
\end{eqnarray}
We have checked that the expressions in \eqref{NLO} reproduce after the Fourier transform \eqref{def:qPDF+pPDF}
the one-loop correction to the qPDFs calculated in \cite{Xiong:2013bka,Stewart:2017tvs}. 

\subsection{Singularities of the Borel transform}\label{sect_renormalons}%
\begin{figure}[b]
\centering
 \includegraphics[width=0.45\textwidth]{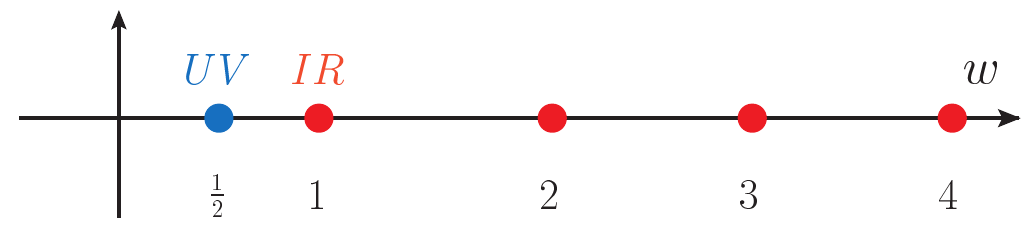}
\caption{Singularity structure of the Borel transform}
\label{fig:borelplane}
\end{figure}

The structure of singularities of the Borel transform of the coefficient functions is illustrated 
in Fig.~\ref{fig:borelplane}. 
There are an UV renormalon singularity at $w=1/2$ and a series of IR renormalons at positive integer $w=1,2,\ldots$.
To the accuracy of our calculation (single bubble chain) all singularities are simple poles.
These singularities  obstruct  the Borel integral in \eqref{BorelIntegral} (Borel-non-summable renormalons) and must be 
matched by the nonperturbative corrections. Note, that the $\widetilde{R}(w)$ term in \eqref{B[H]} is an analytic function of $w$. Thus, it is irrelevant for 
the discussion of singularities.

\subsubsection{Ultraviolet renormalon at $w= 1/2 $}\label{sect_uvren}%
The singularity at $w=1/2$ is generated by the contribution of large momenta in the 
self-energy insertions in the Wilson line and is part of the renormalization factor
\begin{align}
  B[H] &\stackrel{w\to 1/2}{=} \frac{-4C_F}{w-1/2}\sqrt{X}.
\end{align}
This singularity is well-known \cite{Beneke:1994sw} and is in the one-to-one correspondence to the linear UV divergence in the Wilson line's self energy (see also discusion prior to (\ref{n-qITD})). It can be removed by considering  normalized qPDFs, \eqref{n-qITD} or \eqref{hat-qITD},  and will not be considered further in this work.

\subsubsection{Infrared renormalons}\label{sect_irren}%
The  leading IR renormalon singularity is at $w=1$. 
We obtain
\begin{align}
\label{w=1}
B[H^\parallel](w) &\stackrel{w\to 1}{=} \frac{-4C_F}{1-w}\Big[\alpha + \bar\alpha \ln \bar\alpha\Big]X\,,
\notag\\
B[H^\perp](w) &\stackrel{w\to 1}{=} \frac{-4C_F}{1-w}\Big[\alpha + \bar\alpha \ln \bar\alpha + \alpha\bar\alpha\Big]X\,.
\end{align}  
We remind our notation $\bar\alpha = 1-\alpha$.
These expressions present our main result.

Renormalon singularities at $w=n$ ($n=2,3...$) have a generic form
\begin{align}
B[H](w)&=\frac{2C_F}{n-w}\Big[\alpha^np_{n-1}(\alpha)+\frac{(-1)^n\delta(\bar \alpha)}{n!(n\!-\!2)!n^2(2n\!-\!1)}\Big] X^w,
\end{align}
where $p_n(\alpha)$ is a polynomial of order $n$, e.g. $p_1(\alpha)=(5\alpha-3)/6$, $p_2(\alpha)=(\alpha^2-25\alpha+20)/180$, etc.

\section{Power corrections}\label{sect_powercorrections}%

A singularity on the integration path in Eq.~\eqref{BorelIntegral} means that the perturbation theory is 
incomplete and the sum of the series is ill-defined. 
It is customary ~\cite{Beneke:1998ui} to estimate the corresponding ambiguity as
\begin{align}
\delta H(w_0)  = -\pi \frac{1}{\beta_0}  e^{-w_0/(\beta_0 a_s)} \res_{w=w_0}\big[B[H](w)\big]\,, 
\end{align}
where $w_0$ is the position of the singularity and $\displaystyle{\res_{w=w_0}\big[B[H](w)\big]}$ is the corresponding residue. Note that  $ e^{-w_0/(\beta_0 a_s)} = (\Lambda^2/\mu^2)^{w_0}$. Following the standard logic~\cite{Beneke:1998ui,Beneke:2000kc} we assume that this ambiguity must be 
canceled by adding a non-perturbative correction of the same order of magnitude. 

\subsection{Ioffe time quasi-distributions}\label{sect_qITD}%

Considering $\delta H(1)$, we obtain the leading power correction to the qITDs as functions of the ``Ioffe-time''
\begin{align}
 \tau = (p\cdot v) z 
\end{align}
\begin{align}
\label{t4-qITD}
  \mathcal{I}^\parallel(\tau) 
&= I(\tau) +
  \kappa (v^2\! z^2\!\Lambda^2)\! \int_0^1\!d\alpha \, (\alpha + \bar\alpha \ln \bar\alpha)I(\alpha \tau )\,,
\notag\\
  \mathcal{I}^\perp(\tau)
&= I(\tau) +
  \kappa (v^2\! z^2\!\Lambda^2) \!\int_0^1\!d\alpha \, (\alpha + \bar\alpha \ln \bar\alpha + \alpha\bar\alpha)I(\alpha \tau )\,,
\end{align}
where $\kappa$ is a real number of order one. The renormalon ambiguity \eqref{w=1} corresponds to 
\begin{align}
  |\kappa| = -\pi \frac{1}{\beta_0}\left(- \frac{ e^{5/3}}{4 e^{-2\gamma_E}}\right) = \frac{\pi e^{5/3+2\gamma_E}}{4 \beta_0} \simeq 1.5\,,  
\end{align} 
but this number is only indicative. Alternatively, one can put $\kappa=1$ and think of $\Lambda$ as a certain nonperturbative 
parameter of the order of $\Lambda_{\rm QCD}$ that determines an overall normalization of the power correction and 
cannot be fixed in this approach. Note that also the sign of the correction is not determined.

For the normalized qITDs defined in Eq.~\eqref{n-qITD} we obtain instead
\begin{align}
\label{t4-nqITD}
  \mathbf{I}^\parallel(\tau)
&= I(\tau) +
  \kappa (v^2\! z^2\!\Lambda^2) \!\int_0^1\!d\alpha \, [\alpha + \bar\alpha \ln \bar\alpha]_+I(\alpha \tau )\,,
\notag\\
  \mathbf{I}^\perp(\tau)
&= I(\tau) +
  \kappa (v^2\! z^2\!\Lambda^2) \!\int_0^1\!\!d\alpha \, [\alpha + \bar\alpha \ln \bar\alpha + \alpha\bar\alpha]_+I(\alpha \tau )\,,
\end{align} 
where the ``plus'' distribution is defined as usual,
\begin{align}
  [f(\alpha)]_+ = f(\alpha) - \delta(\bar\alpha)\int_0^1 d\beta\, f(\beta)\,. 
\end{align}
The leading power corrections to the qITDs 
$\widehat{\mathbf{I}}^{\parallel(\perp)}$ ~\eqref{hat-qITD} that are normalized to the vacuum correlator,
are the same as for the un-normalized distributions \eqref{t4-qITD}.
The expressions in~\eqref{t4-qITD} and \eqref{t4-nqITD} present the starting point for the following analysis.

In order to visualize the functional dependence of the power correction on the ``Ioffe time'' $\tau$ relative to the leading-twist result, we write 
\begin{align}
    \cI &= I(\tau)\Big\{ 1 + \kappa (v^2 z^2\Lambda^2) \cR_{\cI}(\tau)\Big\},
\notag\\
    \bI &= I(\tau)\Big\{ 1 + \kappa (v^2 z^2\Lambda^2) \bR_{\cI}(\tau)\Big\},
\end{align}
where
\begin{align}
\cR_{\cI}^\parallel(\tau) & = \frac{1}{I(\tau)} \int_0^1\!d\alpha \, 
(\alpha + \bar\alpha \ln \bar\alpha)I(\alpha \tau)\,,
\notag\\  
\cR_{\cI}^\perp(\tau)
& = \frac{1}{I(\tau)} \int_0^1\!d\alpha \, 
(\alpha + \bar\alpha \ln \bar\alpha + \alpha\bar\alpha)I(\alpha \tau)\,.
\end{align}
For the normalized qITDs one obtains 
\begin{align}
  \bR_{\cI}^\parallel(\tau) &= \cR_{\cI}^\parallel(\tau) - \frac14\,, 
\qquad
  \bR_{\cI}^\perp(\tau) &= \cR_{\cI}^\perp(\tau) - \frac5{12}\,, 
\end{align}
and, obviously, 
\begin{eqnarray}
\widehat{\bR}_{\cI}^{\parallel(\perp)}(\tau) = \cR_{\cI}^{\parallel(\perp)}(\tau),
\end{eqnarray}
so that we do not consider them separately.

\begin{figure}[t]
\centering
\includegraphics[width=6.0cm]{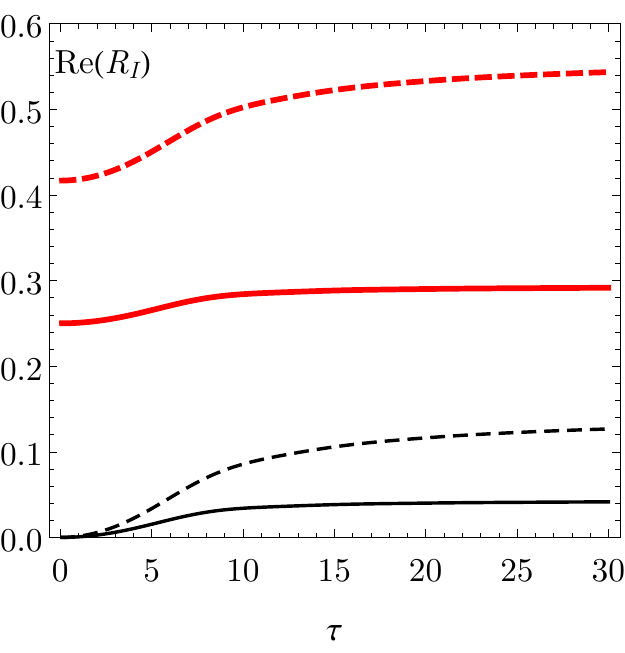}~
\caption{Real parts of $\cR_{\cI}(\tau)$ (thick, red) and $\bR_{\cI}(\tau)$ (thin, black) for the simple quark PDF
model in Eq.~\text{\eqref{q-model}}.
Solid curves are for $\Real R^\parallel$, dashed curves for $\Real R^\perp$.}
\label{fig:R-qITD}
\end{figure}

In general, the qITDs $\cI(\tau)$ and the higher-twist coefficients $\cR_{\cI}(\tau)$ and $\bR_{\cI}(\tau)$ are complex functions, but their imaginary parts appear to be small. The real parts, $\Real\cR_{\cI}(\tau)$ and $\Real\bR_{\cI}(\tau)$,  for a simple model of the valence quark distribution
\begin{align}
\label{q-model}
  q(x) = x^{-1/2}(1-x)^3 
\end{align} 
are plotted in Fig.~\ref{fig:R-qITD}.
The power correction to the ``transverse'' qITD turns out
to be roughly factor four larger than for the ``longitudinal'' qITD.  
In both cases the $R$-functions flatten out at large Ioffe times $\tau \ge 10$ that are, however, hardly accessible in present day lattice
calculations. The normalization to zero proton momentum corresponds to the subtraction of the value at $\tau=0$,
so that for $z\to 0$ there is no power correction by construction. 

\subsection{Parton quasi-distributions}\label{sect_qPDF}%

Making the Fourier transformation of the above results for the qITDs we obtain the qPDFs
\begin{figure*}[t]
\centering
 \includegraphics[width=6.0cm]{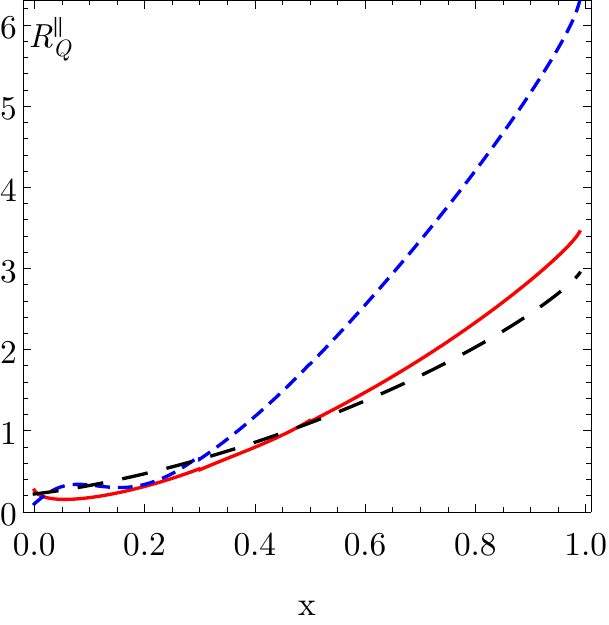}
 ~\includegraphics[width=6.2cm]{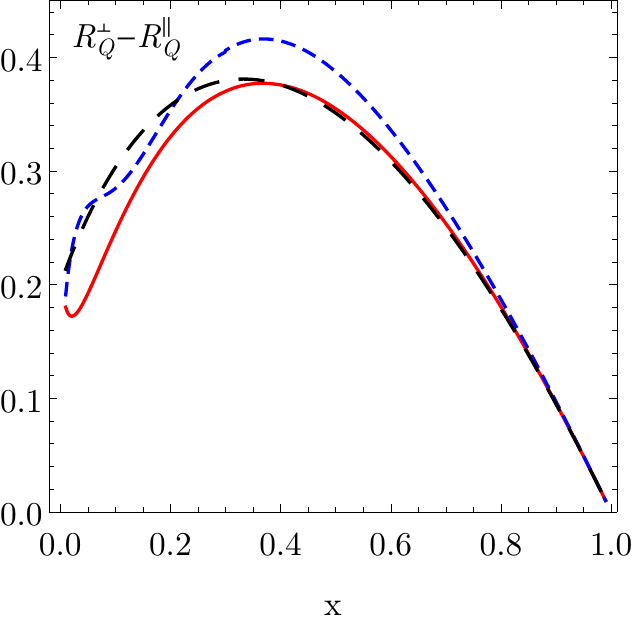}
\caption{$\cR^\parallel_\cQ(x)$ (left panel) and $\cR^\perp_\cQ(x) - \cR^\parallel_\cQ(x)$ (right panel) 
 for MSTW valence u-quarks (long black dashes), d-quarks (short blue dashes) (at 2 GeV), and for the model 
in Eq.~\eqref{q-model} (red solid curves).}
\label{fig:calR}
\end{figure*}
\begin{align}
\label{t4-Qparallel}
 \cQ^\parallel(x,p) &= 
q(x) - \frac{\kappa v^2\! \Lambda^2}{(pv)^2}\Big(\frac{d}{dx} \Big)^2 \int_{|x|}^1\!\frac{dy}{y} \, (y + \bar y \ln \bar y) q(\tfrac{x}{y})
\notag\\&= q(x) - \frac{\kappa v^2 \!\Lambda^2}{x^2(pv)^2} 
\biggl\{
\int_{|x|}^1\!\frac{dy}{y} \frac{y^2}{[1-y]_+} q(\tfrac{x}{y})
\notag\\&\qquad
 + q(x) - |x|q'(x) \biggr\}
\end{align}
and
\begin{align}
\label{t4-Qperp}
 \cQ^\perp(x,p) &= 
q(x) - \frac{\kappa v^2\! \Lambda^2}{(pv)^2}\Big(\frac{d}{dx} \Big)^2\!\!\! \int_{|x|}^1\!\frac{dy}{y}  (y + \bar y \ln \bar y + y\bar y) q(\tfrac{x}{y})
\notag\\&= q(x) - \frac{\kappa v^2 \!\Lambda^2}{x^2(pv)^2} 
\biggl\{
\int_{|x|}^1\!\frac{dy}{y} \left[\frac{y^2}{[1-y]_+} - 2 y^2\right] q(\tfrac{x}{y})
\notag\\&\qquad
 + 2 q(x) - |x|q'(x) \biggr\}
\end{align}
Assuming for what follows $x>0$, these expressions  can be rewritten in the form
\begin{align}
\label{cR}
   \cQ^{\parallel(\perp)}(x,p) &=  q(x) \biggl\{1 - \frac{v^2 \Lambda^2}{ x^2\bar x  (pv)^2}\cR^{\parallel(\perp)}_\cQ(x) \biggr\} 
\end{align}
with
\begin{align}
 R^\parallel_Q(x) &= 
\frac{\bar x }{q(x)} \biggl\{ \int_{x}^1\!\frac{dy}{1-y} \Big[y q(\tfrac{x}{y}) - q(x)\Big] + q(x)-x q'(x)\biggr\}, 
\notag\\
 R^\perp_Q(x) &= 
\frac{\bar x }{q(x)} \biggl\{ \int_{x}^1\!\frac{dy}{1-y} \Big[y (2y-1) q(\tfrac{x}{y}) - q(x)\Big] + 2q(x)  
\notag\\&\qquad - x q'(x) \biggr\}.
\end{align}
Note that we have extracted the prefactor $1/(x^2\bar x)$ for the power correction anticipating that it is 
enhanced as $1/x^2$ and $1/(1-x)$ in the regions of small $x\to 0$ and large $x\to 1$ Bjorken variable, respectively.

The normalized QPDFs $\bQ^{\parallel(\perp)}(x,p)$ are obtained by replacing the kernels in the first lines
in Eqs.~\eqref{t4-Qparallel} and \eqref{t4-Qperp} by the plus distributions, cf.~\eqref{t4-nqITD}.
Writing the result in the form
\begin{align}
\label{bR}
   \bQ^{\parallel(\perp)}(x,p) &=  \hat q(x) \biggl\{1 - \frac{v^2 \Lambda^2}{ x^2\bar x  (pv)^2}\bR^{\parallel(\perp)}_\cQ(x) \biggr\}, 
\end{align}
where $\hat q(x)$ is the quark PDF normalized to the unit integral, one obtains
\begin{align}
 \bR^\parallel_\cQ(x) &= \cR^\parallel_\cQ - \frac14 \frac{\bar x x^2 }{q(x)}  q''(x)\,, 
\notag\\
 \bR^\perp_\cQ(x) &= \cR^\perp_\cQ - \frac5{12} \frac{\bar x x^2 }{q(x)}  q''(x)\,.    
\end{align}
Note that the additional terms are proportional to the second derivative of the quark PDF and thus enhanced as $1/(1-x)^2$ at $x\to 1$.
As already mentioned, the normalization to the vacuum correlator does not affect the $1/p^2$ power corrections so that 
$\widehat{\bR}_{\cQ}^{\parallel(\perp)}(\tau) = \cR_{\cQ}^{\parallel(\perp)}(\tau)$ and we do not need to consider this case separately.

For a numerical study we have used the MSTW NLO valence $u$- and $d$-quark distributions \cite{Martin:2009iq} at the scale 2 GeV and the simple model
in Eq.~\eqref{q-model}. It turns out that the power corrections for (unnormalized) ``longitudinal'' and ``transverse'' qPDFs 
are similar in size so that we show $\cR^\parallel_\cQ(x)$ on the left panel in Fig.~\ref{fig:calR}, and the difference 
$\cR^\perp_\cQ(x) - \cR^\parallel_\cQ(x)$ on the right panel.

The $x$-dependence of $\cR_\cQ(x)$ is similar for all quark PDF models: The power correction is small for $x\to 0$ 
(but non-zero, $\cR^\parallel_\cQ(0)\sim 0.2$, $\cR^\perp_\cQ(0)\sim 0.4$) and 
increases steeply with $x$ (almost linearly). For the $u$-quark, the result is very similar to the simple model in Eq.~\eqref{q-model} whereas for the 
$d$-quark the power correction is, roughly, factor two larger. Note that the difference $\cR^\perp_\cQ(x) - \cR^\parallel_\cQ(x)$ depends on the 
PDF model only very weakly. 

\begin{figure*}[tbp]
\centering
 \includegraphics[width=5.0cm]{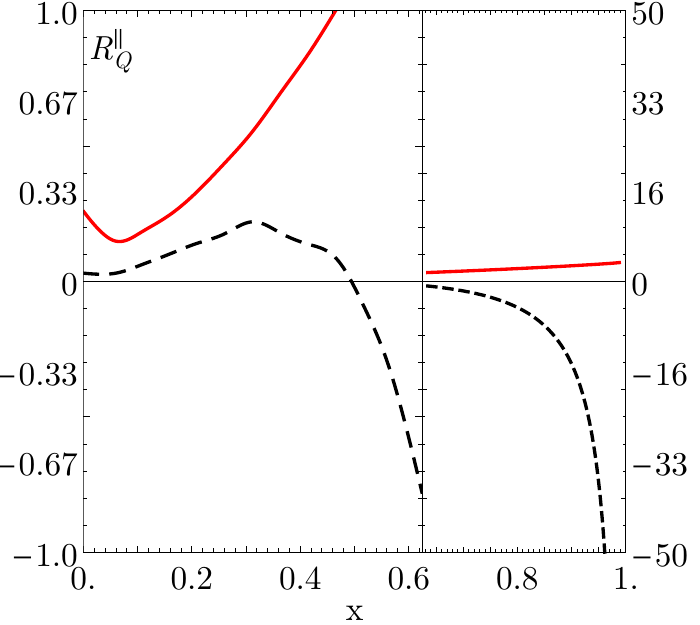}~
 \includegraphics[width=5.0cm]{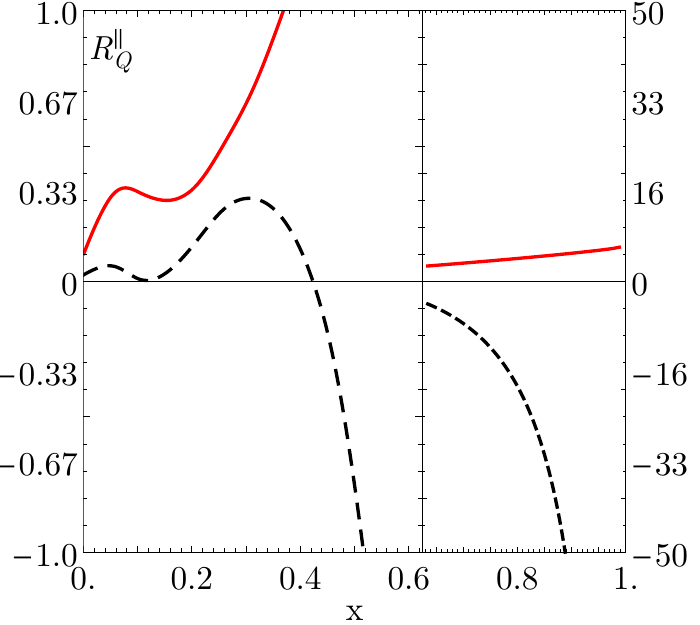}~
 \includegraphics[width=5.0cm]{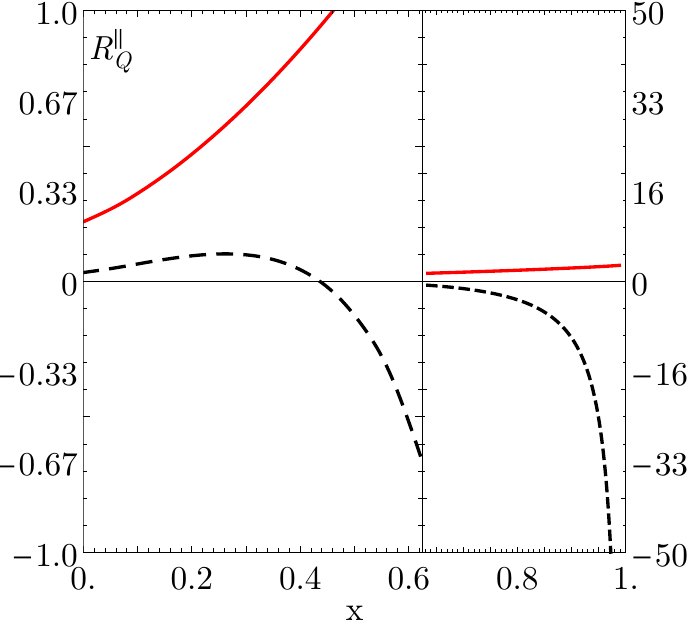}
\\
 \includegraphics[width=5.0cm]{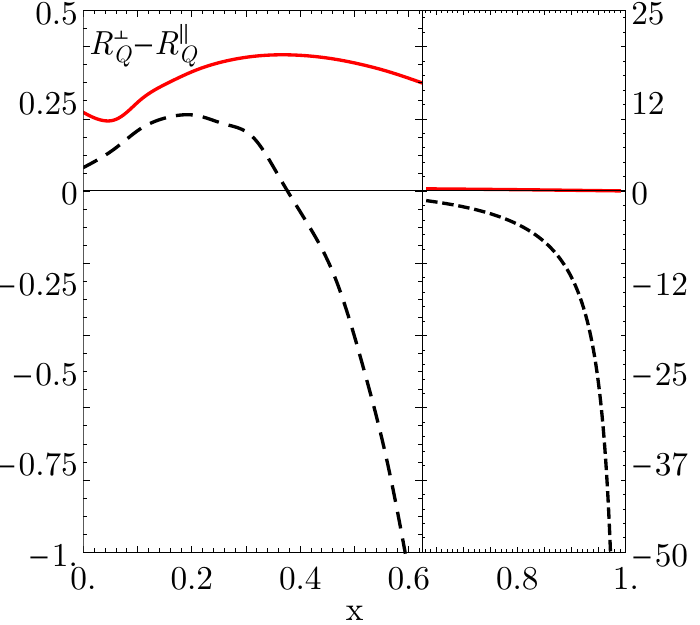}~
 \includegraphics[width=5.0cm]{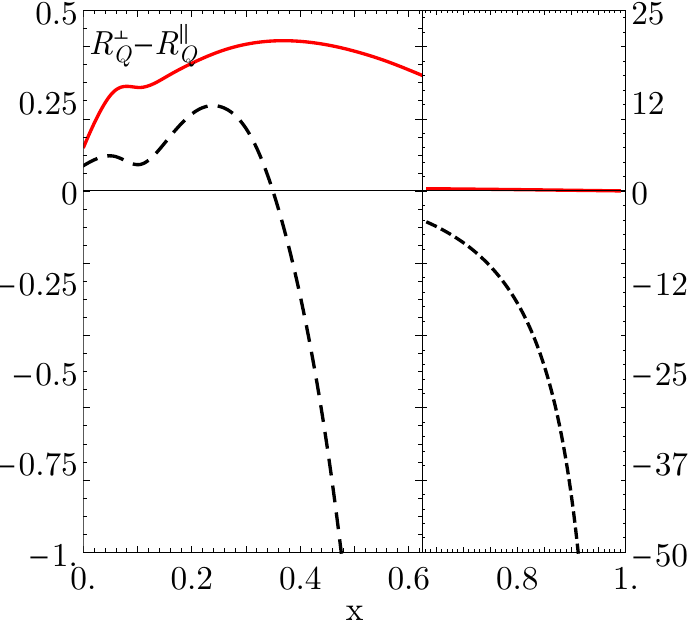}~
 \includegraphics[width=5.0cm]{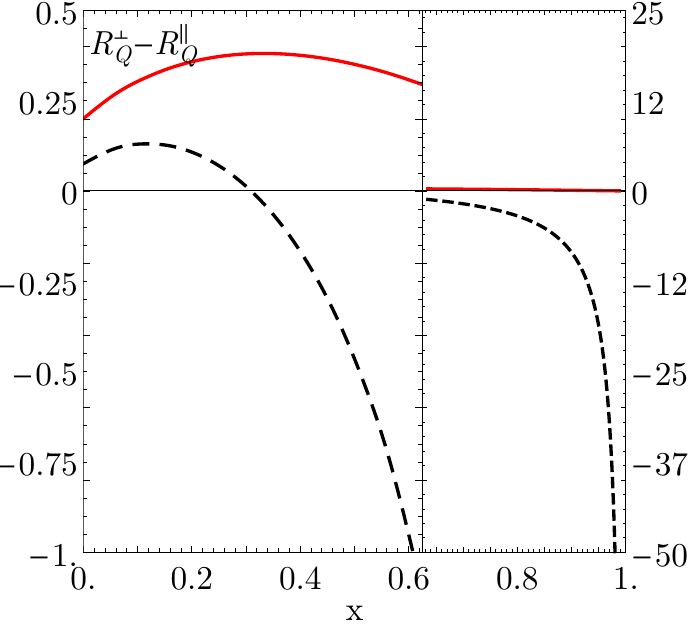}
\caption{Upper panels:
$\cR^\parallel_\cQ$ (solid red curves)  and $\bR^\parallel_\cQ$ (dashed black curves) 
for MSTW valence u-quarks (left), d-quarks (middle), both at 2 GeV, and for the  model 
in Eq.~\eqref{q-model} (right).
Lower panels: the same for 
$\cR^{\perp}_\cQ-\cR^{\parallel}_\cQ $ and $\bR^{\perp}_\cQ-\bR^{\parallel}_\cQ $.
Note split panels: the results for $0<x<0.6$ and $0.6<x<1$ are shown using a different scale on the vertical axis.}
\label{fig:bRparallel}
\end{figure*}

Constructing the qPDFs from the qITDs normalized to the value at zero momentum has a large effect.
This is illustrated  in Fig.~\ref{fig:bRparallel} where in the upper panels we show the results for the ``longitudinal'' case and in 
the lower panels the difference between the ``longitudinal'' and ``transverse'' distributions.  
The three panels (from left to right) correspond to the MSTW valence u-quarks, d-quarks, and the simple model 
in Eq.~\eqref{q-model}, respectively. The red solid lines stand for  $\cR_\cQ(x)$ 
(i.e. the same as in Fig.~\ref{fig:calR}), and the result for the normalized qPDFs, $\bR_\cQ(x)$, is shown by the black dashed curves.

We see that the normalization to the qITD at zero momentum significantly reduces the power correction at moderate values of $x \lesssim 0.5$
at the cost of dramatical increase at higher $x$ values. This normalization procedure is, therefore, not suitable to access the large-$x$ behavior
of the PDFs, but apparently minimizes power corrections in the not-so-large $x$ region. 

\subsection{Parton pseudo-distributions}\label{sect_pPDF}%

\begin{figure}[t]
\centering
 \includegraphics[width=5.0cm]{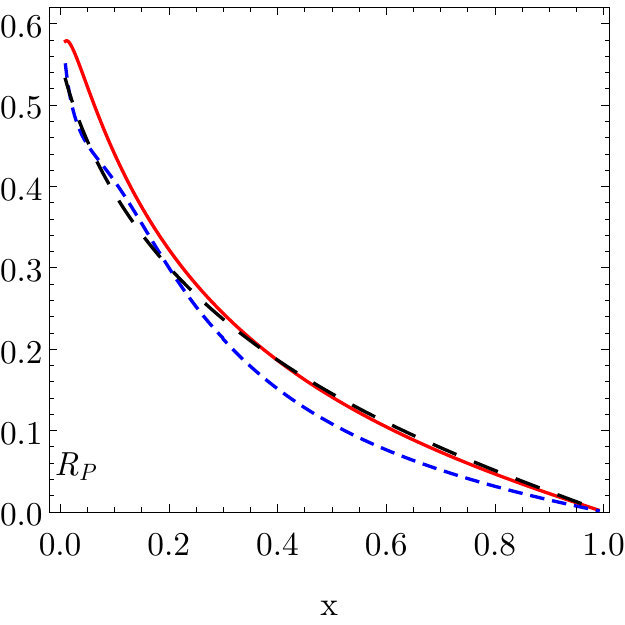}~
\caption{\label{fig:pPDF} Power correction to the pPDF $\mathcal{R}_\cP$ \eqref{t4-pPDF}
for MSTW valence  u-quarks (long black dashes), d-quarks (short blue dashes), both at 2 GeV, and for the simple model 
$q(x) = x^{-1/2} (1-x)^3 $ (red solid curve).}
\end{figure}

Power corrections for the pPDF can be obtained easily from the corresponding expression for the 
``transverse'' qITDs \eqref{t4-qITD}, \eqref{t4-nqITD}. 
Writing the result as
\begin{align}
\mathcal{P}(x,z,\mu) &= q(x)\Big\{1 + (v^2\! z^2\!\Lambda^2)  \theta(|x|<1) \mathcal{R}_\cP(x)\Big\},
\end{align}
and similarly for the pPDF normalized to zero momentum, $\mathbf{P}(x,z)$, we obtain
\begin{align}
\label{t4-pPDF}
\mathcal{R}_\cP(x) &= \frac{1}{q(x)} \int_{|x|}^1\!\frac{dy}{y} \, (y + \bar y \ln \bar y +  y \bar y ) q(\tfrac{x}{y})\,,
\notag\\
\mathbf{R}_\cP(x) &= \frac{1}{q(x)} \int_{|x|}^1\!\frac{dy}{y} \, [y + \bar y \ln \bar y +  y \bar y ]_+ q(\tfrac{x}{y}) 
\notag\\&= 
\mathcal{R}_\cP(x) - 5/12 \,.
\end{align}
The numerical results are shown in Fig.~\ref{fig:pPDF}.

Note that $\mathcal{R}_\cP(x)$ is very similar for all considered models for the valence quark PDFs and
decreases at $x\to 1$. Indeed, it is easy to see that $\mathcal{R}_\cP(x) = \mathcal{O}(1-x)$ in this limit, 
similar to the target mass correction, Eq.~\eqref{target-pPDF}. This suppression is removed once pPDF is
normalized to zero momentum ~\cite{Orginos:2017kos} (which adds a negative constant), but can be upheld if
normalized to the vacuum expectation value of the  same operator. 

\section{Conclusions}

We have presented an analysis of power-suppressed (higher-twist) contributions to qPDFs and pPDFs based on the study of factorial divergences (renormalons) in the corresponding coefficient functions within the bubble-chain approximation. Factorial asymptotic implies that the sum of the series is only defined to a power accuracy and therefore, the QCD perturbation theory must be corrected by nonperturbative power-suppressed contributions to produce unambiguous predictions. Our results have to be considered as a ``minimal model'' for the higher-twist corrections that captures effects that are necessary for the selfconsistency of the theory, but possibly misses other nonperturbative corrections that are, e.g, protected 
by symmetries and not ``seen'' through perturbative expansions. Our main conclusions are as follows:
\begin{itemize}
\item Position space PDFs (qITDs) have flat power corrections at large Ioffe times. Generally, power corrections are much 
      larger for the ``transverse'' projection as compared to the ``longitudinal'' projection. 
\item Power corrections for qPDFs have a generic behavior
\begin{align}
  \cQ (x,p) &=  q(x) \biggl\{1 + \mathcal{O}\left( \frac{\Lambda^2}{p^2}\cdot\frac{1}{{x^2(1-x)}}\right) \biggr\}.     
\end{align}
Note that the corresponding target mass corrections, \eqref{target-parallel} and \eqref{target-perp}, do not show up the  $1/x^2$ enhancement. This behavior is commensurate to the suppression of Nachtmann's target mass corrections $\sim x^2 m^2/Q^2$ at small $x$ for the DIS structure functions. The normalization of the underlying qITDs to unity at the zero momentum considerably reduces the size the power correction to the qPDFs at $ x \lesssim 0.5$ at the cost of a strong enhancement at $x\gtrsim 0.5-0.6$. Thus, such a normalization procedure is not suitable to access the large-$x$ behavior of the PDFs, but apparently minimizes power corrections in the intermediate $x$-region. Note that the above discussion is based on a normalization procedure different from the nonperturbative renormalization used in ref.~\cite{Chen:2018xof}, thus the power corrections for the latter might behave differently from what has been shown here.
 \item
Power corrections for pPDFs have a generic behavior
\begin{align}
\mathcal{P}(x,z) &= q(x)\Big\{1 + \mathcal{O}\big( z^2\!\Lambda^2{(1-x)}\big)\Big\}.
\end{align}
The suppression at $x\to 1$ is lifted by the zero-momentum normalization factor. However, it can be upheld by the normalization to the vacuum matrix element of the same operator. We conclude that pPDFs can offer an interesting alternative to qPDFs for the study of large-$x$ region.
\end{itemize}
As a byproduct of this study, we provide the results for the coefficient functions 
in the large-$n_f$ approximation, terms $\sim n_f^k\alpha_s^{k+1}$,  to all orders in perturbation theory.
These results can be useful to estimate the effects of uncalculated higher orders and scale-setting 
using the BLM-type procedure. The corresponding analysis goes beyond the scope of this paper and will be presented in a future publication.
 
\section*{Acknowledgements}

This  work  was  partially  supported  by  the  DFG grant SFB/TRR-55 ``Hadron Physics from Lattice QCD''.


\appendix 


%
\section{Leading-twist coefficient function in the bubble-chain approximation}
\label{App:H} 

In this appendix we present some details of the calculation of coefficient function $H$ in the bubble-chain approximation. The calculation is performed by explicit evaluation and renormalization of diagrams with $n$-insertions of bubbles. The evaluation closely follows the one presented in details in refs.\cite{Beneke:1995pq,Scimemi:2016ffw}, with the main difference being that it is made directly in the coordinate space.

To obtain the twist-two coefficient function in $\MSbar$-scheme for the operator
\begin{eqnarray}
O_v^\mu(z,0)=\bar q(z v)\gamma^\mu [zv,0]q(0),
\end{eqnarray}
it is sufficient to calculate its free-quark matrix element (with massless on-shell quarks). Since there is only a single scale $z^2$, the expression for diagrams have simple structure that contains poles in $\epsilon$ and $\ln(z^2\mu^2)$. The poles in $\epsilon$ correspond to UV poles (to be renomalized) and collinear poles (to be incorporated into the definition of collinear PDF). Therefore, the coefficient function is given by the finite part of the (renormalized) diagrams (for a more detailed discussion see \cite{Echevarria:2016scs,Scimemi:2016ffw}).

The diagrams contributing to the large-$n_f$ limit are presented in Fig.\ref{fig:bubblechain}. 
The (bare) propagator with the insertion of $n$-fermion loops in the coordinate representation reads
\begin{eqnarray}
\Delta_n^{\mu\nu}&=&-\frac{1}{8\pi^{d/2}}R_\epsilon^n\frac{\Gamma(2-(n+1)\epsilon)}{\Gamma(2+n\epsilon)}
\\\nonumber &&
\times\Big[\frac{1+2n\epsilon}{1-(n+1)\epsilon}g^{\mu\nu}+2\frac{x^\mu x^\nu}{x^2}\Big]\frac{1}{(-x^2+i0)^{1-(n+1)\epsilon}},
\end{eqnarray}
where $d=4-2\epsilon$ is the dimension of space-time in the dimension regularization, 
\begin{eqnarray}
R_\epsilon=-\frac{2n_f}{3}\frac{a_s}{\epsilon}\frac{6\Gamma(1+\epsilon)\Gamma^2(2-\epsilon)}{4^\epsilon\Gamma(4-2\epsilon)},
\end{eqnarray}
and $a_s=g^2/(4\pi)^{d/2}$. One can check explicitly that this propagator satisfies the integral equation
\begin{eqnarray}
\Delta_{n}^{\mu\nu}(x)=\int d^dz \Delta^{\mu\rho}_{n-m}(x-z)\Delta^{\rho\nu}_m(z),
\end{eqnarray}
with $1\leqslant m \leqslant n-1$. This equation reflects the fact that the composition of two bubble-chains is again a bubble-chain. The $\epsilon$-poles in $\Delta_n$ correspond to UV poles of fermion loops, and to be renormalized later.

The leading large-$n_f$ diagrams structurally reproduce the one-loop diagrams calculated in the Landau gauge. Due to the gauge invariance, the transverse part of the gluon propagator cancels in the sum of diagrams, and therefore, the complete result can be obtained by the use of only longitudinal part, which significantly simplifies the calculation. The reduced propagator reads
\begin{eqnarray}
\tilde \Delta_n^{\mu\nu}&=&\frac{-1}{4\pi^{d/2}}
\frac{\Gamma(1-(n+1)\epsilon)}{\Gamma(1+n\epsilon)}\frac{g^{\mu\nu} R_\epsilon^n}{(-x^2+i0)^{1-(n+1)\epsilon}}.
\end{eqnarray}
The expressions for diagrams evaluated with this propagator are (notation corresponds to Fig.\ref{fig:bubblechain})
\begin{widetext}
\begin{eqnarray}
(a)^\mu &=&2a_sC_F \frac{R_\epsilon^n}{4^\epsilon}(-v^2z^2)^{(n+1)\epsilon}\frac{\Gamma(-(n+1)\epsilon)}{\Gamma(2+n\epsilon)}
\\&& \nonumber \hfill
\times\Big\{\int_0^1  
(O_v^\mu(z,\alpha z)-O_v^\mu(z,0))(1-\epsilon)\bar \alpha^{1+n\epsilon}\,_2F_1(1,2-(n+2)\epsilon,2+n\epsilon;\bar \alpha)~d\alpha-\frac{1}{2}O^\mu_n(z,0)\Big\},
\\
(b)^\mu &=&2a_sC_F \frac{R_\epsilon^n}{4^\epsilon}(-v^2z^2)^{(n+1)\epsilon}\frac{\Gamma(-(n+1)\epsilon)}{\Gamma(2+n\epsilon)}
\\&& \nonumber \hfill
\times\Big\{\int_0^1  
(O_v^\mu(\bar \alpha z,0)-O_v^\mu(z,0))(1-\epsilon)\bar \alpha^{1+n\epsilon}\,_2F_1(1,2-(n+2)\epsilon,2+n\epsilon;\bar \alpha)~d\alpha-\frac{1}{2}O^\mu_n(z,0)\Big\},
\\
(c)^\mu &=&2a_sC_F \frac{R_\epsilon^n}{4^\epsilon}(-v^2z^2)^{(n+1)\epsilon}\frac{\Gamma(-(n+1)\epsilon)}{\Gamma(1+n\epsilon)}
\\&& \nonumber \hfill
\times\int_0^1  d\alpha \int_0^{\bar \alpha} d\beta (1-\alpha-\beta)^{n\epsilon}\Big[(1+n\epsilon)O_v^\mu(\bar \alpha z,\beta z)-2(n+1)\epsilon \frac{v^\mu v_\nu}{v^2}O^\nu_v(\bar \alpha z,\beta z)\Big],
\\
(d)^\mu &=&-2a_sC_F \frac{R_\epsilon^n}{4^\epsilon}(-v^2z^2)^{(n+1)\epsilon}\frac{\Gamma(-(n+1)\epsilon}{\Gamma(1+n\epsilon)}\frac{O^\mu_v(z,0)}{1-2(n+1)\epsilon}.
\end{eqnarray}
The renormalization of these diagrams is straightforward and discussed in details e.g. in ref.\cite{Beneke:1995pq}. Next, we present the diagram-by-diagram renormalized expression for the coefficient function. One obtains
\begin{eqnarray}\label{app:a}
(a)^\mu_{\text{ren}}&=&\frac{2C_F}{\beta_0}(\beta_0 a_s)^{n+1}\biggl\{\int_0^1 \frac{d\alpha dt}{t} (O_v^\mu(z,t\alpha z)-O_v^\mu(z,0))\Big[
\frac{r_{n+1}(\bar \alpha)+r_n(\bar \alpha)}{n+1}+(-1)^nn!g_0^{[n+1]}\big(\bar \alpha t^2 \mathbb{z}^2\big)\Big]
\\\nonumber && 
\hfill-\frac{1}{2}O_v^\mu(z,0)\Big[\frac{h_{n+1}^{[0]}}{n+1}+(-1)^nn!h_0^{[n+1]}\big(\mathbb{z}^2\big)\biggr\},
\\
(b)^\mu_{\text{ren}}&=&\frac{2C_F}{\beta_0}(\beta_0 a_s)^{n+1}\biggl\{\int_0^1\frac{d\alpha dt}{t}  (O_v^\mu((1-t\alpha) z,0)-O_v^\mu(z,0))\Big[
\frac{r_{n+1}(\bar \alpha)+r_n(\bar \alpha)}{n+1}+(-1)^nn!g_0^{[n+1]}\big(\bar \alpha t^2\mathbb{z}^2\big)\Big]
\\\nonumber && 
\hfill-\frac{1}{2}O_v^\mu(z,0)\Big[\frac{h_{n+1}^{[0]}}{n+1}+(-1)^nn!h_0^{[n+1]}\big(\mathbb{z}^2\big)\biggr\},
\\
(c)^\mu_{\text{ren}}&=&-\frac{2C_F}{\beta_0}(\beta_0 a_s)^{n+1}\int_0^1 d\alpha \int_0^{\bar \alpha} d\beta\biggl\{\Big[\frac{r_{n+1}(\gamma)+2r_n(\gamma)+r_{n-1}(\gamma)}{n+1}
\\\nonumber && +(-1)^nn!g_0^{[n+1]}(\gamma \mathbb{z}^2)-(-1)^nn!g_0^{[n]}(\gamma \mathbb{z}^2)\Big]O_v^\mu(\bar \alpha z,\beta z)
+2(-1)^n n!g_0^{[n]}(\gamma \mathbb{z}^2)\frac{v^\mu v_\nu}{v^2}O_v^\nu(\bar \alpha z,\beta z)\biggr\},
\\\label{app:d}
(d)^\mu_{\text{ren}}&=&\frac{2C_F}{\beta_0}(\beta_0 a_s)^{n+1}\Big(\frac{y_{n+1}^{[0]}}{n+1}+(-1)^nn!y_0^{[n+1]}(\mathbb{z}^2)\Big)O_n^{\mu}(z,0),
\end{eqnarray}
\end{widetext}
where $\mathbb{z}^2=-z^2 v^2\mu^2/4$, $\gamma=1-\alpha-\beta$. In these expressions we have promoted the factors $(-2n_f/3)$ to the $\beta_0$ coefficient of QCD, $\beta_0=11/3C_A-2n_f/3$. The functions $g_j^{[n]}$, $h_j^{[n]}$, etc., are defined as coefficients of expansion
\begin{eqnarray}
G(\epsilon,s,\mathbb{z}^2)&=&\sum_{k,j=0}^\infty g_n^{[j]}(\mathbb{z}^2)s^j \epsilon^n,
\\
\frac{G(\epsilon,s,\mathbb{z}^2)}{1-s+\epsilon}&=&\sum_{k,j=0}^\infty h_n^{[j]}(\mathbb{z}^2)s^j \epsilon^n,
\\
\frac{G(\epsilon,s,\mathbb{z}^2)}{1+2s}&=&\sum_{k,j=0}^\infty y_n^{[j]}(\mathbb{z}^2)s^j \epsilon^n,
\\
x^\epsilon G(\epsilon,0,\mathbb{z}^2)&=&\sum_{k=0}^\infty r_n(x) \epsilon^n,
\end{eqnarray}
where
\begin{eqnarray}
G(\epsilon,s,\mathbb{z}^2)&=&\Big(\mathbb{z}^2\Big)^{-s}\frac{\Gamma(1+s)}{\Gamma(1-s+\epsilon)}
\\\nonumber && \times \Bigg(\frac{6 \Gamma(1-\epsilon)\Gamma^2(2+\epsilon)}{\Gamma(4+2\epsilon)}\Bigg)^{s/\epsilon-1}.
\end{eqnarray}
Note, that operators on the r.h.s. of (\ref{app:a}-\ref{app:d}) are renormalized operators. Taking the hadronic matrix element of these diagrams and performing the Borel summation we obtain the final result given in Eq.\,(\ref{B[H]}).

The factorial divergences  $\sim \beta_0^n  n!$ that correspond to Borel non-summable renormalons 
can be inferred directly from the expressions in (\ref{app:a}-\ref{app:d}) by inspection. 
These terms give rize to the renormalon contribution in (\ref{B[H]}). The remaining terms are regular and 
contribute to the function $\widetilde R(w)$ in Eq.\,(\ref{B[H]}). 
Finally, specializing to the particular $n$ values one obtains the $\sim a_s^{n+1}n_f^n$ contributions to the coefficient function. 
At $n=0$ one obtains the NLO coefficient function, quoted in (\ref{NLO}). 
The $n=1$ result corresponds to the $n_f$ part of the NNLO coefficient function, Eq.\,(\ref{NNLO}).

%

\bibliographystyle{apsrev}
\bibliography{BVZbibliography}%

\end{document}